\documentclass[11pt]{article}

\usepackage[utf8]{inputenc}
\usepackage[T1]{fontenc}

\usepackage{anysize}
\marginsize{2.1cm}{2.1cm}{1.60cm}{1.60cm}

\usepackage{hyperref}
\hypersetup{colorlinks=true,filecolor=blue,linkcolor=blue,citecolor=blue,urlcolor=blue,pdfnewwindow=true}
\usepackage{amssymb}
\usepackage{amsmath}
\usepackage{amsthm}
\usepackage{xcolor}
\usepackage{graphicx}
\usepackage{rotating}
\usepackage{algorithm}
\usepackage{multirow}
\usepackage{multicol}
\usepackage{algorithmic}
\usepackage{graphicx}
\usepackage{enumerate}
\usepackage{url}
\usepackage{hhline}
\usepackage{subcaption}
\usepackage[round]{natbib}
\usepackage{natbib}
 \bibpunct[, ]{(}{)}{,}{a}{}{,}
 \def\bibfont{\small}
 \def\bibsep{\smallskipamount}
 \def\bibhang{24pt}

\newcommand{\D}{\displaystyle}

\newcommand{\T}{^\mathrm{T}}

\allowdisplaybreaks

\begin{document}

\title{A joint model for author's characteristics and collaboration pattern in bibliometric networks: a Bayesian approach}

\author{Stefano Nasini\thanks{Corresponding author: IESE Business School, University of Navarra, Barcelona, Spain, SNasini@iese.edu} \and V\'ictor Mart\'inez-de-Alb\'eniz\thanks{IESE Business School, University of Navarra, Barcelona, Spain, valbeniz@iese.edu. V. Mart\'{i}nez-de-Alb\'{e}niz's research was supported in part by the European Research Council - ref. ERC-2011-StG 283300-REACTOPS and by the Spanish Ministry of Economics and Competitiveness (Ministerio de Econom\'{\i}a y Competitividad) - ref. ECO2011-29536.}\and Tahereh Dehdarirad\thanks{ Department of Library and Information Science, University of Barcelona, Barcelona, Spain, tdehdari@gmail.com} }

\maketitle

\begin{abstract}

Demographic and behavioral characteristics of journal authors are important indicators of \emph{homophily} in co-authorship networks. In the presence of correlations between adjacent nodes (assortative mixing), combining the estimation of the individual characteristics and the network structure results in a well-fitting model, which is capable to provide a deep understanding of the linkage between individual and social properties. This paper aims to propose a novel probabilistic model for the joint distribution of nodal properties (authors' demographic and behavioral characteristics) and network structure (co-authorship connections), based on the nodal similarity effect. A Bayesian approach is used to estimate the model parameters, providing insights about the probabilistic properties of the observed data set. After a detailed analysis of the proposed statistical methodology, we illustrate our approach with an empirical analysis of co-authorship of 1007 journal articles indexed in the ISI Web of Science database in the field of neuroscience between 2009 and 2013.

\end{abstract}

\bigskip


\section{Introduction}\
\label{Section:Introduction}
The increasing specialization of scientific research, the interdisciplinary character of most projects, and the increased funding of cross-institution initiatives have made scientists take part in collaboration networks \citep{EMBR:EMBR2011161, Haeussler2013688}. Co-authorship networks represent a widely studied class of collaboration networks. They have been extensively studied using different statistical approaches \citep{Newman06042004, Newman-assort-2003}, with the purpose of identifying the structure of scientific partnerships and the role played by the individual researchers.

The majority of methodological contributions focus on modelling the structure of scientific co-authorship, based on the projection of a two-mode network (author--paper network) into a one-mode structure of co-authorship (author--author network), where links represent co-authors, i.e., authors sharing common papers, as described by \cite{Leydesdorff2008317}. We use a similar approach here. Our main contribution is to combine individual and social properties of journal authors, by internalizing the effect of authors' similarities in their patterns of connection. It provides an insight into the level of \emph{homophily} in co-authorship networks, in terms of specific socio-demographic characteristics \citep{Newman-assort-2003}, while accounting for relevant network features based on observed nodal properties.

The notation we use denotes $\mathcal{V}$ the set of $N$ authors and $\mathcal{E} \subseteq \mathcal{V} \times \mathcal{V}$ their known structure of connections; $\mathcal{K}$ denotes a set of $K$ of categorical properties (in our application, $\mathcal{K} = \{genders, \, nationalities\}$) defined for each author in $\mathcal{V}$. The nodal similarities are assumed to reflect the overlap of authors' categorical statuses, with respect to the properties in $\mathcal{K}$.

An exponential random model is proposed to internalize the effect of nodal similarities on the joint distribution of network and authors' properties \citep{caimo2011bayesian, Robins2007192}. Exponential families possess good properties that typically simplify the statistical inference of model parameters. As we explain in Section \ref{Section:model}, the inclusion of nodal similarities as sufficient statistics for the exponential random model entails the impossibility of a complete characterization of the probability distribution, due to the intractability of the normalizing constant. This represents one of the strongest barriers to the numerical-optimization of the likelihood-function and legitimates the use of simulation-based approaches -- such as the Monte Carlo maximum likelihood of \cite{GeyerThompson1992} and pseudo-likelihood estimation of \cite{StraussIkeda1990}.

As suggested by \citealt{caimo2011bayesian}, this drawback can be overcome by embedding the defined model into a Bayesian estimation framework, which reformulate the estimation problem based on the ability of simulating from the posterior distribution. We build on \citealt{murray2006}, which proposed a MCMC method to simulate from this class of distributions, allowing a flexible estimation of the effect of nodal similarity -- which is the main scope of this paper. Our approach is able to generate the following insights:
\begin{itemize}
\item \small \emph{we can estimate author' collaborations based on their demographic and behavioral characteristics;}
\item \small \emph{we can estimate author' demographic and behavioral characteristics based on their pattern of connections.}
\end{itemize}
In other words, our model connects nodal properties with network structure, so one can be used to predict the other. We illustrate our method through the analysis of co-authorship of over a thousand journal articles between 2009 and 2013 in the neuroscience research community.

Section \ref{Section:DataSet} introduces and describes the co-authorship data set, along with the relevant network statistics we aim to control in a probabilistic model. Section \ref{Section:model} provides a detailed description of the proposed exponential random model for this type of data set and embeds such model in a general Bayesian framework. Section \ref{Section:estimation} takes into account the algorithmic aspects of the estimation of the model parameters. The numerical results are presented in Section \ref{Section:NumericalResults}. Section \ref{Section:Conclusion} concludes.

\section{Presenting the co-authorship data set}\
\label{Section:DataSet}
The data set is composed of the scientific publications indexed in the Web of Science (WOS) database between 2009 and 2013 in the field of Neuroscience. 153,182 research papers were retrieved in the first step. Then, we conducted stratified random sampling. The sample size was determined with a $3\%$ sampling error and $95\%$ of level of confidence. Table \ref{tab:SampleDescription} shows the total number of publications and the stratified sample size per year in the studied field.

\begin{table}[H]
\begin{center}
\scalebox{0.82}{
\begin{tabular}{|r|cc|}
\hline
Year            & $\#$ publications ($\%$)   & Stratified sample size       \\
\hline
2009            & 28,819 (18.81$\%$)                    & 199                   \\
2010            & 30,154(19.69$\%$)                      & 208                   \\
2011            & 31,030 (20.26$\%$)                     & 214                   \\
2012            & 31,265 (20.41$\%$)                     & 218                   \\
2013            & 31,914 (20.83$\%$)                     & 221                   \\
\hline
Total           & 153,182                           & 1,060                 \\
\hline
\end{tabular}}
\caption{\label{tab:SampleDescription} \footnotesize The total number of publications and the stratified sample size, 2009-2013.}
\end{center}
\end{table}

\begin{figure}[H]
        \centering
        \begin{subfigure}[b]{0.32\textwidth}
                \includegraphics[width=\textwidth]{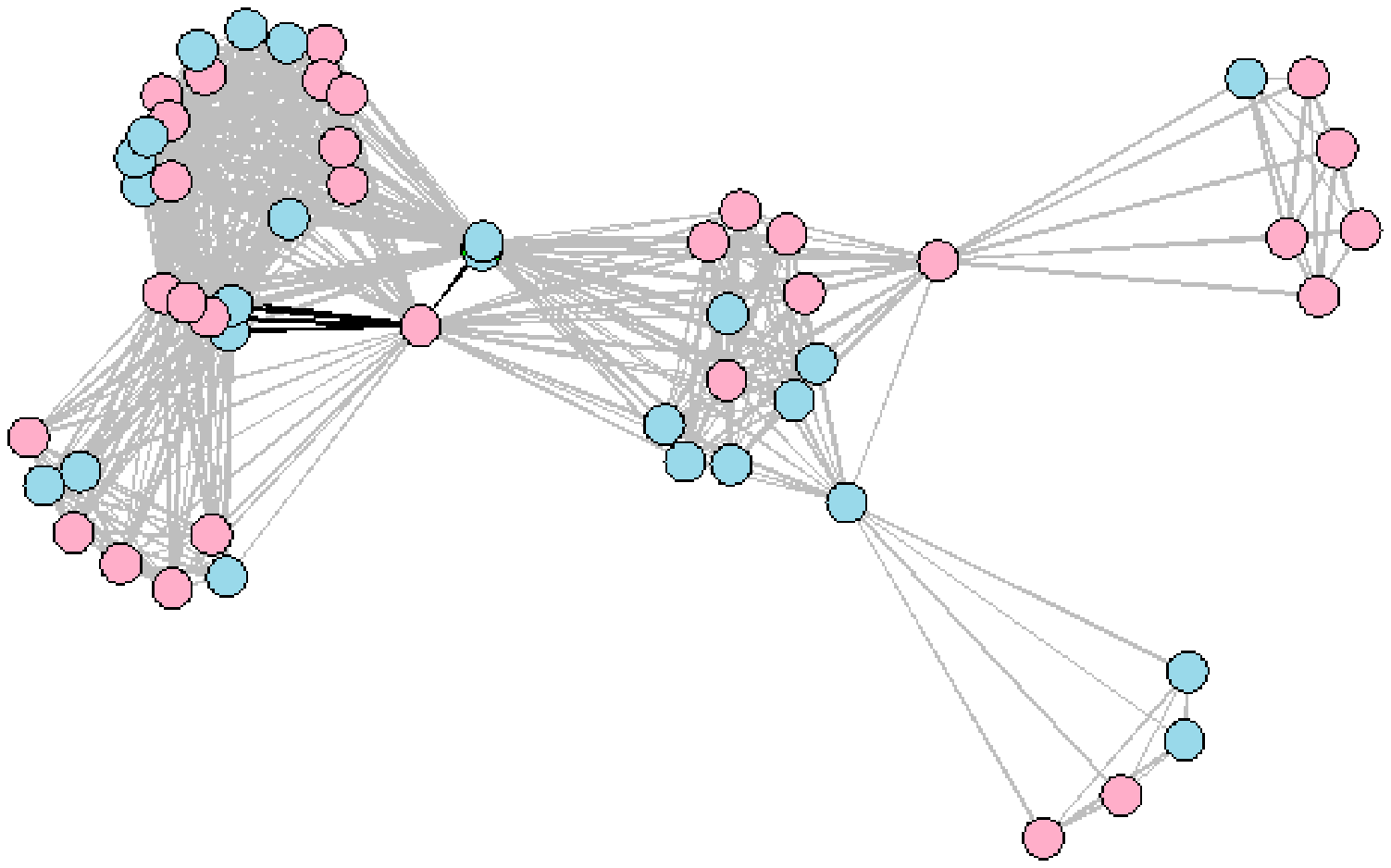}
                \caption{\scriptsize First largest component.}
        \end{subfigure}%
        ~
        \begin{subfigure}[b]{0.32\textwidth}
                \includegraphics[width=\textwidth]{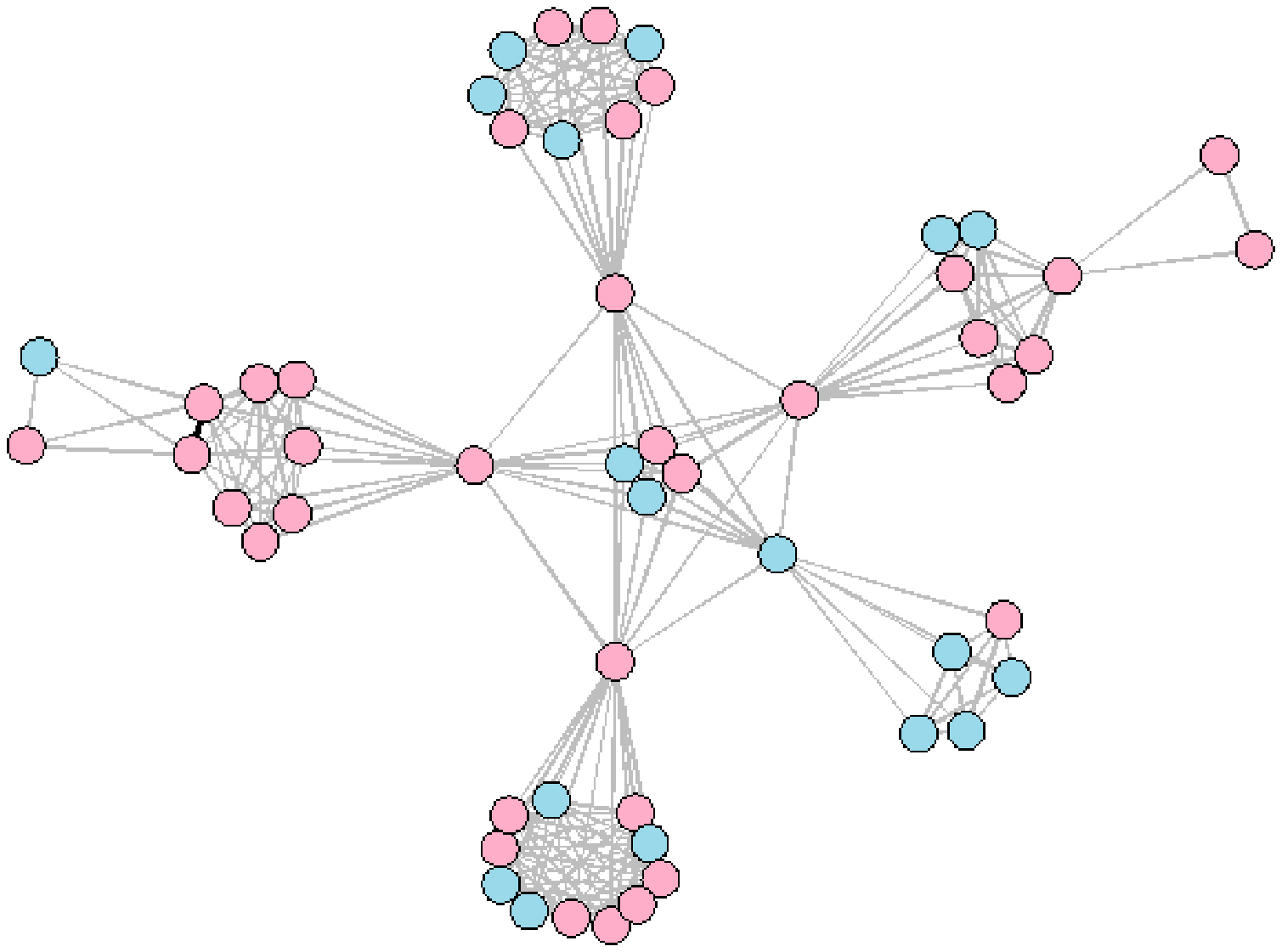}
                \caption{\scriptsize Second largest component.}
        \end{subfigure}
        ~
        \begin{subfigure}[b]{0.32\textwidth}
                \includegraphics[width=\textwidth]{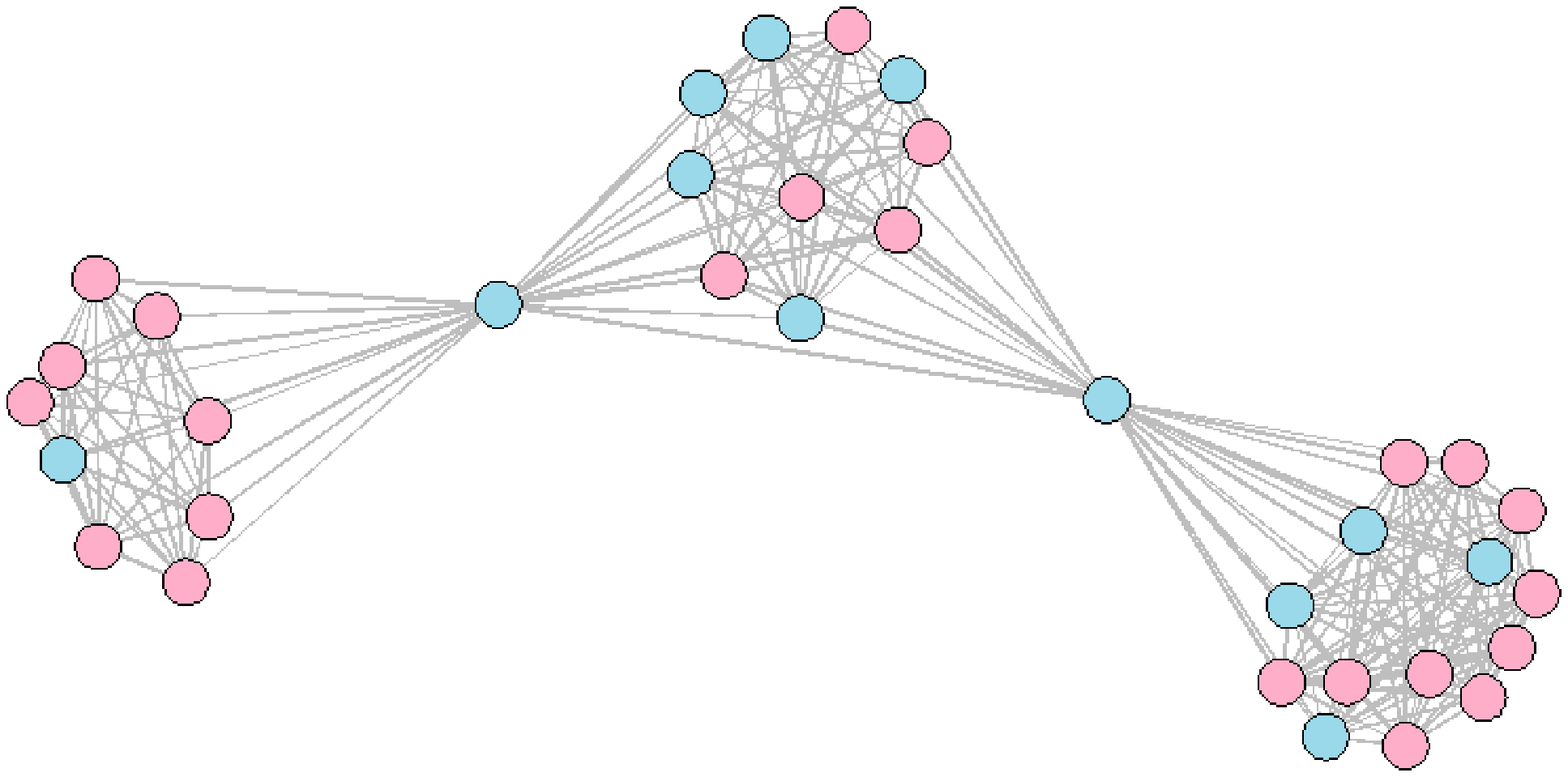}
                \caption{\scriptsize Third largest component.}
        \end{subfigure}
\caption{\footnotesize The three largest components with nodal genders (blue for men, pink for women).}\label{fig:ConnectedComponents1}
\end{figure}

\begin{figure}[H]
        \centering
        \begin{subfigure}[b]{0.32\textwidth}
                \includegraphics[width=\textwidth]{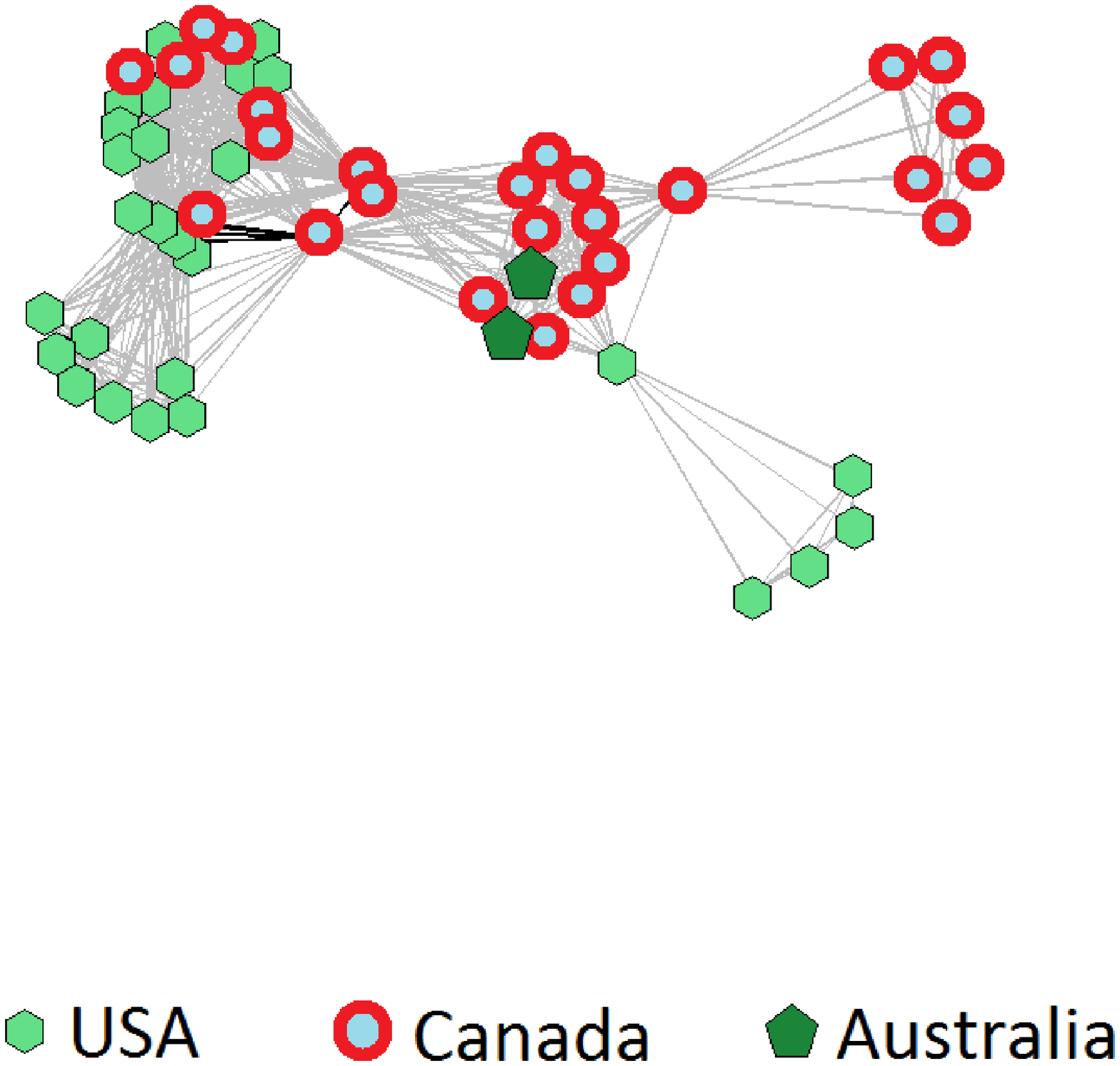}
                \caption{\scriptsize First largest component.}
        \end{subfigure}%
        ~
        \begin{subfigure}[b]{0.32\textwidth}
                \includegraphics[width=\textwidth]{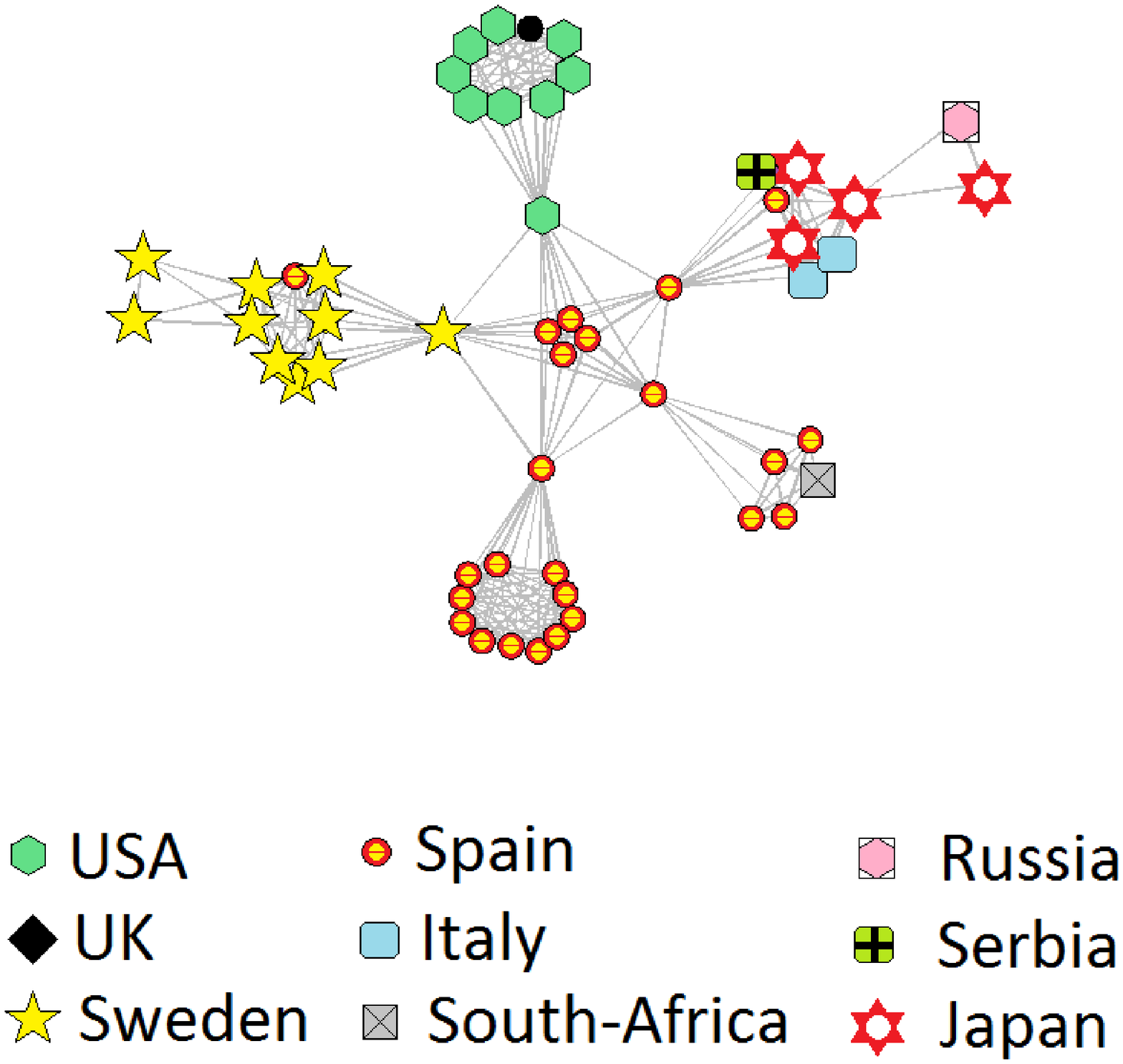}
                \caption{\scriptsize Second largest component.}
        \end{subfigure}
        ~
        \begin{subfigure}[b]{0.32\textwidth}
                \includegraphics[width=\textwidth]{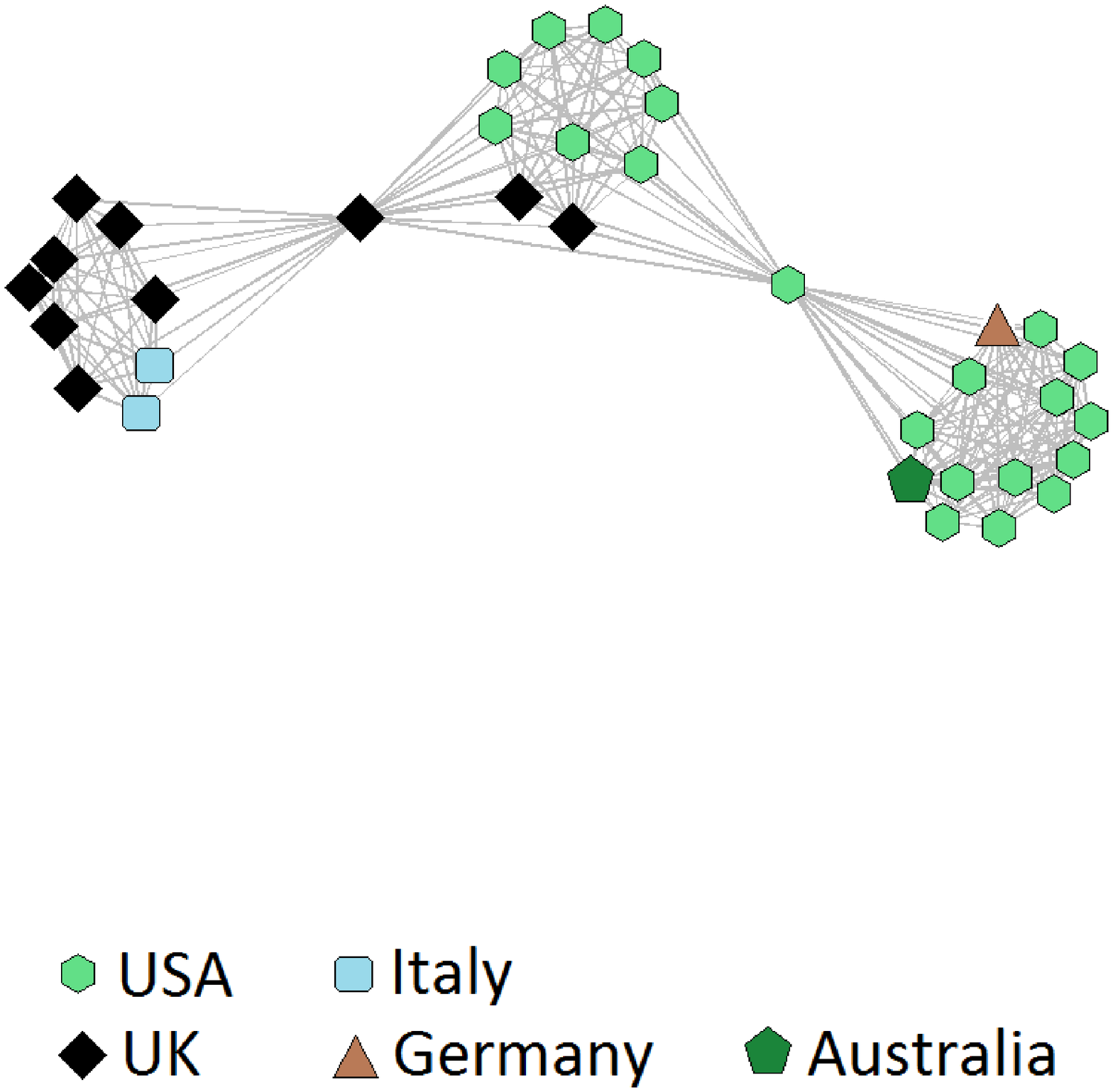}
                \caption{\scriptsize Third largest component.}
        \end{subfigure}
\caption{\footnotesize The three largest components with nodal nationalities.}\label{fig:ConnectedComponents2}
\end{figure}

Finally, after eliminating those papers whose gender was unclear (53 of them, $5\%$), our data set comprised 1,007 ($95\%$) of 1,060 papers. These $1,007$ papers were used as our data set for further analysis, corresponding to $5,385$ authors. Each author was assigned to a nationality based on the authors' country of affiliation. Thus, for each author, two demographic characteristics were collected: gender and nationality.

A network structure of scientific collaboration between authors was then generated by connecting those authors whose names jointly appear in one or more of the $1,007$ articles. The resulting network comprised $207$ disconnected components and the three largest had respectively size of $55$, $53$ and $35$, as shown in Figures \ref{fig:ConnectedComponents1} and \ref{fig:ConnectedComponents2}, associated with the corresponding nodal gender and nationalities.

\section{Model definition and specification}
\begin{figure}[H]
        \centering
        \includegraphics[scale = 0.28]{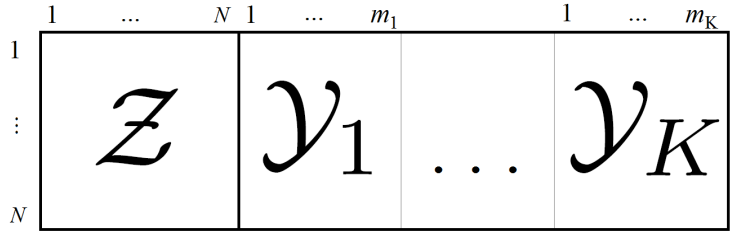}
\caption{\footnotesize Sample space.}\label{fig:SampleSpace}
\end{figure}

\
\label{Section:model}
This section proposes an exponential random model which internalizes the structure of dependencies of individual characteristics and network structure. Consider a categorical variable $\mathbf{Y}_k$, with $m_k$ categories defined on a set of $N$ individuals, and its representation in term of an $N \times m_k$ binary matrix $\mathbf{y}_k \in \{0,1\}^{m_k n}$. Let $\mathcal{Y}_k$ be the set of all possible realizations of $\mathbf{Y}_k$. Similarly, let $\mathbf{Z}$ be the adjacency matrix of a random network with $N$ nodes and $\mathcal{Z} \subseteq \{0,1\}^{N \times N}$ the set of its possible realizations. The sample space under consideration can be defined as $\mathcal{X} = \mathcal{Z} \times \mathcal{Y}_1 \times \ldots \times \mathcal{Y}_K \subseteq \{0,1\}^{N \times N + \sum_{k=1}^{K}m_k}$ -- the set of network structures among $N$ individuals, taking $K$ categorical properties, as illustrated in Figure \ref{fig:SampleSpace}.

Let $\mathbf{X}$ be a random matrix, taking values in $\mathcal{X}$ and $\mathbf{x}$ a possible realization. In the exponential family of distributions the conditional probability of $\mathbf{x} \in \mathcal{X}$ takes the following form: $P(\mathbf{x} ~|~ \theta) \propto \exp(T(\mathbf{x})\T \boldsymbol{\theta})$, where $\boldsymbol{\theta}$ is a vector of \emph{natural parameters} of the distribution, which can usually take any value in the reals; and $T(\mathbf{x})$ is a vector of sufficient statistics.

The exponential random model developed here internalizes the nodal similarity effect by including the matching in the categorical properties of each pair of nodes as a sufficient statistic of the exponential distribution:
\begin{equation}\label{eq:JointProb}
P(\mathbf{x} ~|~ \boldsymbol{\alpha}, \boldsymbol{\beta}, \boldsymbol{\gamma} )  \propto \left\{
 \begin{array}{l l}
    \displaystyle   \exp \left[ \boldsymbol{\alpha} \T A(\mathbf{y}) \, + \, \boldsymbol{\beta} \T B(\mathbf{z}) \, + \, \boldsymbol{\gamma}\T G(\mathbf{y}, \mathbf{z}) \right] & \quad \text{if $\mathbf{x} \in \mathcal{X}$}\\
    0 & \quad \text{otherwise}
\end{array} \right.
\end{equation}
where $\mathbf{y}$ is a matrix whose $(h_k, r)$ component represents the dummy indicator of whether an individual $r$ has value $h_k$ in the $k^{th}$ categorical variable, for $h_k = 1 \ldots m_k$, $r \in \mathcal{V}$; $\mathbf{z}$ is the adjacency matrix whose $(r,s)$ component represents the binary indicator of the existence of a connection between $r \in \mathcal{V}$ and  $s \in \mathcal{V}$. $A(\mathbf{y})$ is a vector of sufficient statistics which accounts for combinatorial properties of the categorical variables $\mathbf{y}$ (such as the number of nodes per each level of each categorical variable, number of associated categories, etc.) only, and $\boldsymbol{\alpha}$ is the corresponding vector parameter. Similarly, $B(\mathbf{z})$ is a vector of sufficient statistics which accounts for combinatorial properties of the network structure $\mathbf{z}$ (such as the clustering coefficient, the assortativity coefficient, the average path length, etc.), but independent of nodal exogenous properties; and $\boldsymbol{\beta}$ is the corresponding vector parameter. The interaction between nodal characteristics and connections variables is internalized into the model by the sufficient statistics $G(\mathbf{y}, \mathbf{z})$; $\boldsymbol{\gamma}$ is the parameter vector associated with these interactions.

The specification of the sample space $\mathcal{X}$ can incorporate both network and nodal properties, in accordance with our modeling assumptions and our need to control specified combinatorial properties \citep{Castro&Nasini:2015}. In other words, $P(\mathbf{x} ~|~ \boldsymbol{\alpha}, \boldsymbol{\beta}, \boldsymbol{\gamma} )  = 0$ if $\mathbf{x}$ does not satisfy a set of feasibility constraints. As illustration, three possible sample spaces are shown in \eqref{mpmcurn:1} by exogenously fixing the degree sequence, the number of edges and the size of each categorical level. They are specified in term of the solution sets of systems of linear constraints. Note that intersections of these sets give rise to hybrid sample spaces with complex combinatorial structures.

\begin{equation*} \label{mpmcurn:1}
\begin{array}{ll}

\mbox{ fixed degree sequence } \qquad &  \begin{array}{|lll}
\D \sum_{h_k = 1}^{m_k} y_{h_k r} = 1                    & k = 1 \ldots K, \, r = 1 \ldots n  \\
\D \sum_{r=1}^{n} z_{rs} = d_{s}                           & r \in \mathcal{V}
\end{array}  \\
& \\
\mbox{ fixed number of edges } \qquad & \begin{array}{|lll}
\D \sum_{h_k = 1}^{m_k} y_{h_k r} = 1                              & k = 1 \ldots K, \, r = 1 \ldots n \\
\D \sum_{(r, \, s) \in \mathcal{V} \times \mathcal{V} } z_{rs} = d  &
\end{array} \\
& \\
\mbox{ fixed categorical levels } \qquad & \begin{array}{|lll}
\D \sum_{h_k = 1}^{m_k} y_{h_k r} = 1                             & k = 1 \ldots K, \, r = 1 \ldots n \\
\D \sum_{r = 1}^{m} y_{h_k r} = f_k                               & k = 1 \ldots K,
\end{array}\\
& \\
\end{array}
\end{equation*}

Classical inferential methods for the model parameters of \eqref{eq:JointProb} are encumbered by the intractability of the normalizing constant, which makes the numerical optimization of the likelihood function very challenging. The next section describes an MCMC algorithm to simulate from the Bayesian posterior distribution of the parameter.

\section{Estimation method}\
\label{Section:estimation}
As noted by \cite{murray2006} and by \cite{caimo2011bayesian}, the intractability of the normalizing constants of most random network models entails a ``double intractability" of the posterior distribution when the model is embedded in a Bayesian framework. This is also true for model \eqref{eq:JointProb}. MCMC algorithms are often used to draw samples from distributions with intractable normalization constants. However, they do not apply to a doubly-intractable constant.

Consider the kernel of the probability function \eqref{eq:JointProb} and let $\mathbf{x}^{(0)} \in \mathcal{X}$ be the observed data set --the co-authorship network structure $\mathbf{z}^{(0)}$, the nodal genders $\mathbf{y}^{(0)}_1$, the nodal nationalities $\mathbf{y}^{(0)}_2$. Given a prior distribution $\pi(\boldsymbol{\alpha}, \boldsymbol{\beta}, \boldsymbol{\gamma})$, apply the Bayes rule:
\begin{equation*}\label{eq:posterior}
\qquad P(\boldsymbol{\alpha}, \boldsymbol{\beta}, \boldsymbol{\gamma} ~|~ \mathbf{x}^{(0)}) = \frac{P(\mathbf{x}^{(0)} ~|~ \boldsymbol{\alpha}, \boldsymbol{\beta}, \boldsymbol{\gamma} ) \pi(\boldsymbol{\alpha}, \boldsymbol{\beta}, \boldsymbol{\gamma}) }{\displaystyle \int_{\alpha,\beta,\gamma} P(\mathbf{x}^{(0)} ~|~ \boldsymbol{\alpha}, \boldsymbol{\beta}, \boldsymbol{\gamma} ) \pi(\boldsymbol{\alpha}, \boldsymbol{\beta}, \boldsymbol{\gamma}) \, d \boldsymbol{\alpha} \, d \boldsymbol{\beta} \, d \boldsymbol{\gamma}}
\end{equation*}

Since both $P(\mathbf{x}^{(0)} ~|~ \boldsymbol{\alpha}, \boldsymbol{\beta}, \boldsymbol{\gamma} )$ and $P(\boldsymbol{\alpha}, \boldsymbol{\beta}, \boldsymbol{\gamma} ~|~ \mathbf{x}^{(0)})$ can only be specified under proportionality conditions, \cite{murray2006} proposed a MCMC approach which overcomes the drawback to a large extent, based on the simulation of the joint distribution of the parameter and the sample spaces, conditioned to the observed data set $\mathbf{x}_0$, that is to say, $P(\mathbf{x}, \, \boldsymbol{\alpha}, \, \boldsymbol{\beta}, \, \boldsymbol{\gamma} ~|~ \mathbf{x}_0)$. We follow the same approach. Our application of the Metropolis-Hastings method \citep{Bolstad2010} to simulate from such distribution is summarized in Algorithm \ref{alg:MH_ExchangeAlgorithm}.

\begin{algorithm}[H]
\begin{algorithmic}[1]
{ \small
\STATE Initialize $(\boldsymbol{\alpha}, \, \boldsymbol{\beta}, \, \boldsymbol{\gamma})$
\REPEAT
    \STATE Draw $(\boldsymbol{\alpha}^\prime, \, \boldsymbol{\beta}^\prime, \, \boldsymbol{\gamma}^\prime)$ from $h( . ~|~ \boldsymbol{\alpha}, \, \boldsymbol{\beta}, \, \boldsymbol{\gamma})$
    \STATE Draw $\mathbf{x}^{\prime}$ from $P( . ~|~ \boldsymbol{\alpha}^\prime, \, \boldsymbol{\beta}^\prime, \, \boldsymbol{\gamma}^\prime)$
    \STATE Accept $(\boldsymbol{\alpha}^\prime, \, \boldsymbol{\beta}^\prime, \, \boldsymbol{\gamma}^\prime)$ with probability $\min \left \{ 1, \displaystyle \frac{P(\mathbf{x}^{\prime} ~|~ \boldsymbol{\alpha}, \boldsymbol{\beta}, \boldsymbol{\gamma} ) P(\mathbf{x}^{(0)}  ~|~ \boldsymbol{\alpha}^\prime, \boldsymbol{\beta}^\prime, \boldsymbol{\gamma}^\prime ) \pi(\boldsymbol{\alpha}^\prime, \boldsymbol{\beta}^\prime, \boldsymbol{\gamma}^\prime) }{ P(\mathbf{x}^{(0)} ~|~ \boldsymbol{\alpha}, \boldsymbol{\beta}, \boldsymbol{\gamma} ) P(\mathbf{x}^{\prime} ~|~ \boldsymbol{\alpha}^\prime, \boldsymbol{\beta}^\prime, \boldsymbol{\gamma}^\prime) \pi(\boldsymbol{\alpha}, \boldsymbol{\beta}, \boldsymbol{\gamma})} \right\}$
    \STATE Update $(\boldsymbol{\alpha}, \, \boldsymbol{\beta}, \, \boldsymbol{\gamma})$
\UNTIL{Convergence}}
\end{algorithmic}
\caption{\label{alg:MH_ExchangeAlgorithm} \footnotesize Exchange algorithm of \cite{murray2006}.}
\end{algorithm}
The distribution $h()$ is used to simulate candidate points from the posterior and it is here assumed to be symmetric.
Note that in step 3 of Algorithm \ref{alg:MH_ExchangeAlgorithm} a new value of the parameters $(\boldsymbol{\alpha}^\prime, \, \boldsymbol{\beta}^\prime, \, \boldsymbol{\gamma}^\prime)$ is randomly proposed and in step 4 a sample from $\mathcal{X}$ is simulated with probability given in \eqref{eq:JointProb}. Clearly, this is a computationally intensive procedure.

\section{Numerical results and analysis}\
\label{Section:NumericalResults}
In this section the described probabilistic model is applied to the second largest component of the co-authorship network in the neuroscience community, as shown in figures \ref{fig:ConnectedComponents1} and \ref{fig:ConnectedComponents2}, where $n = 54$ (nodes), $K = 2$ (categorical properties), $m_1 = 2$ (genders), $m_2 = 9$ (nationalities). The sample space is defined as the Cartesian product between the set of $n$ node undirected networks with fixed number of edges $d=234$ and the set of all possible realization of $2$ categorical variables with $2$ and $9$ levels. The following specification of model \eqref{eq:JointProb} is taken into account in this section
\begin{equation} \label{eq:JointProb_Spec}
P(\mathbf{x})  \propto \left\{
 \begin{array}{l l}
    \displaystyle   \exp \left[ \sum_{k \in \mathcal{K}} \sum_{h_k=1}^{m_k} \alpha_{k, h_k} \left(\sum_{r \in \mathcal{V}} y_{h_k r}\right) + \sum_{k \in \mathcal{K}} \gamma_k \sum_{(r, \, s) \in \mathcal{V} \times \mathcal{V} } z_{rs} \left( \sum_{h_k = 1}^{m_k} y_{h_k r}y_{h_k s}\right) \right] & \quad \text{if $\mathbf{x} \in \mathcal{X}$}\\
    0 & \quad \text{otherwise}
\end{array} \right.
\end{equation}
The vector of sufficient statistics thus contains the total amount of nodes for each property and each level, and the association between edges and two nodal properties
The corresponding natural parameters are $\alpha_{k, h_k}$ and $\gamma_{k}$.

It is important to keep in mind which interpretation should be given to the estimated natural parameters $\alpha_{k,h_k}$ and $\gamma_{k}$, for $h_k = 1 \ldots m_k$, $k \in \mathcal{K}$. In the case of uniform distribution within the sample space $\mathcal{X}$, we should have $\alpha_{k,h_k} = 0$; hence deviations in $\alpha_{k,h_k} \neq 0$ can be interpreted as incorporating different proportions of the properties within the nodal population (e.g., nationalities not being evenly present in the sample). Moreover, if nodal and network properties are independent, then $\gamma_{k} = 0$: the association between connections and nodal similarities has no effect on the probability of observing a given configuration in $\mathcal{X}$. Any non-zero value of the natural parameters entails a deviation from such independence. Specifically, if $\gamma_{k} = 0$, \eqref{eq:JointProb_Spec} is reduced to the product between the probability mass functions of $K$ multinomial random variables $\text{Multinom}(e^{\alpha_{k,1}} \ldots e^{\alpha_{k,m_k}})$ and the Erdos-Reniy random Graph model with fixed number of edges $d$. In the numerical results we set $\gamma_{k} = 0$ as a null model for the computational comparison.

\begin{figure}[H]
        \centering
        \begin{subfigure}[b]{0.45\textwidth}
                \includegraphics[width=\textwidth]{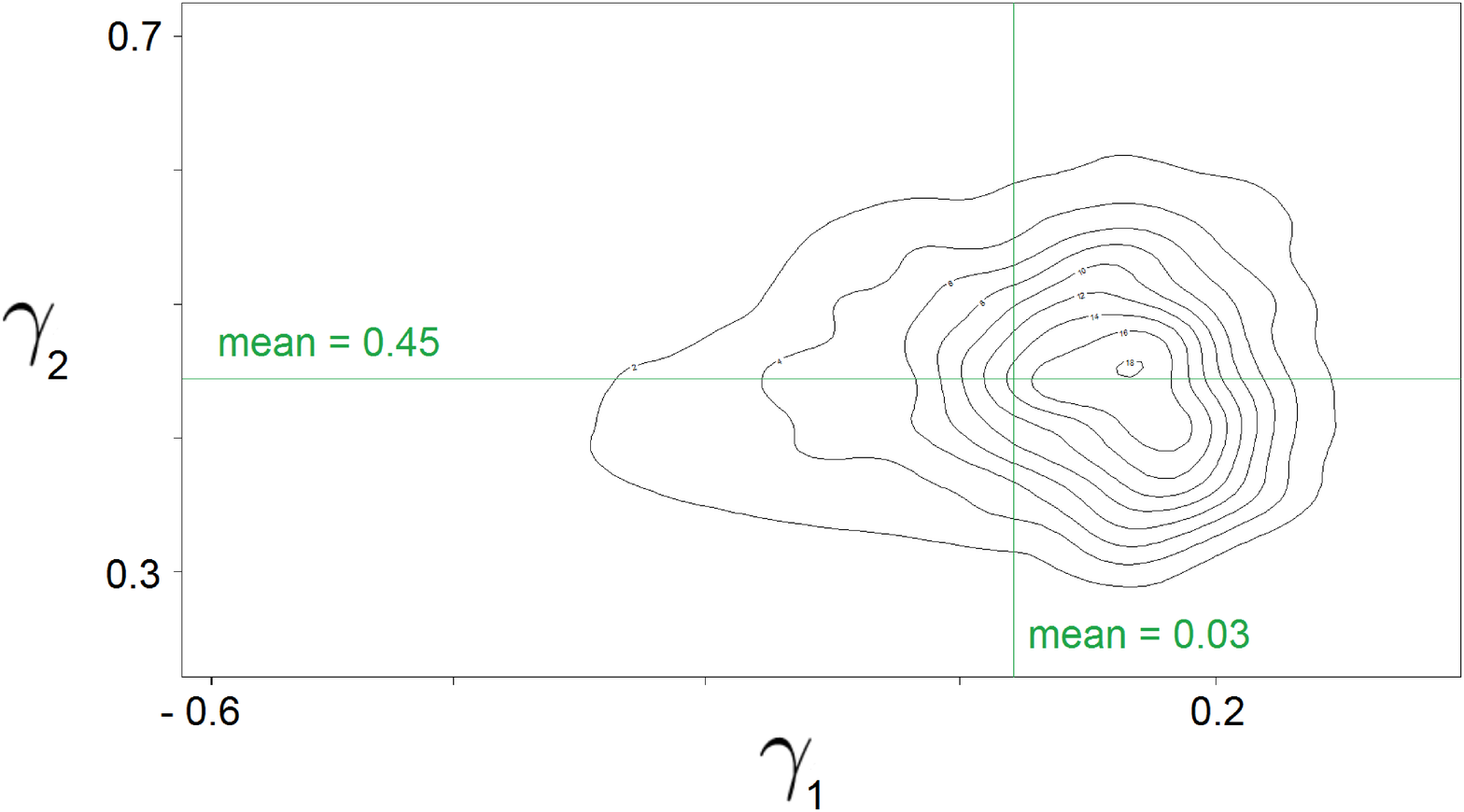}
                \caption{\scriptsize Contour plot of the empirical posterior}
        \end{subfigure}%
        ~
        \begin{subfigure}[b]{0.37\textwidth}
                \includegraphics[width=\textwidth]{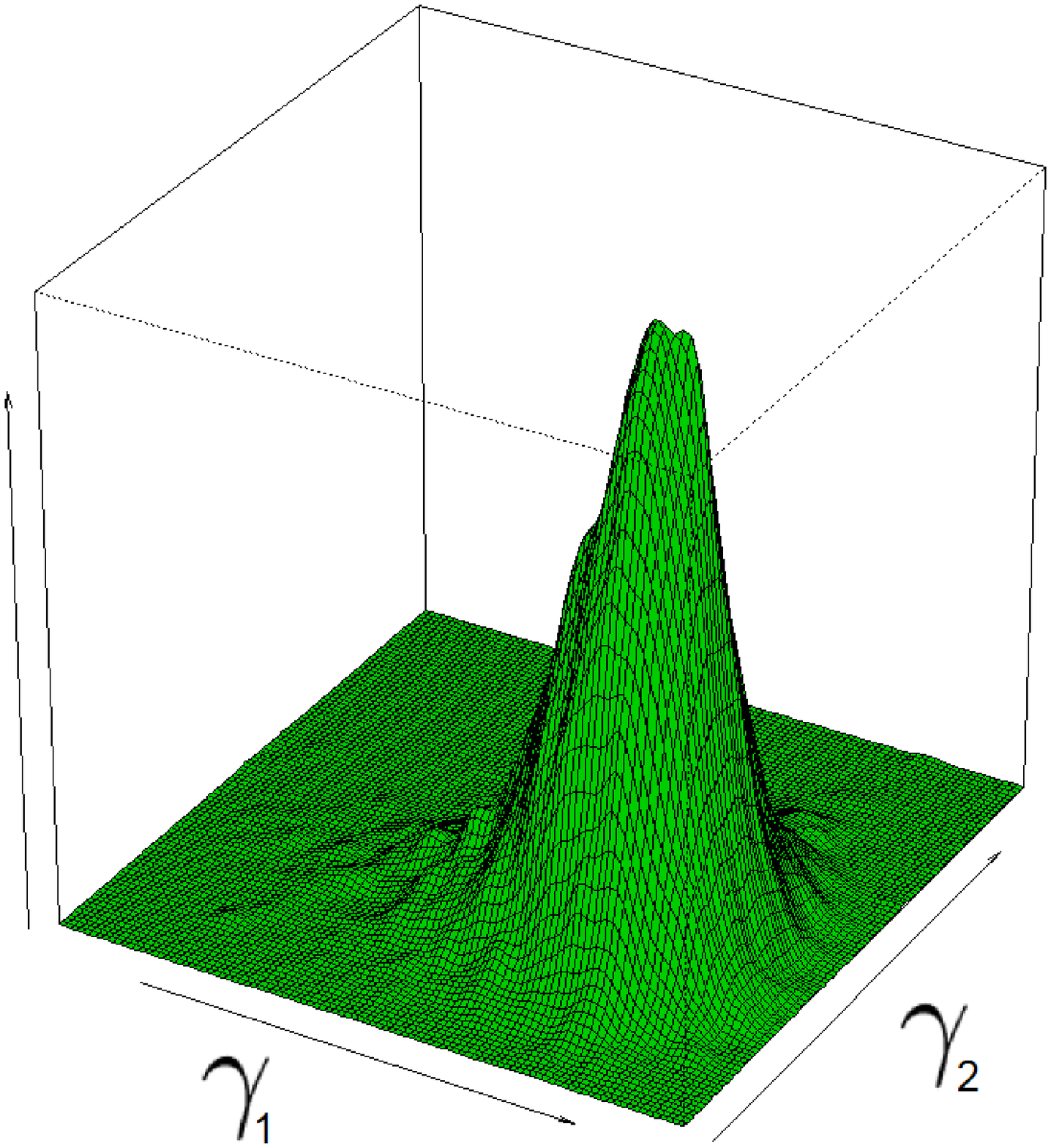}
                \caption{\scriptsize Bidimensional empirical posterior}
        \end{subfigure}
\caption{\footnotesize Marginal posterior of $(\gamma_{1}, \gamma_{2})$, corresponding to the second largest component of the co-authorship data set.}\label{fig:ContourGamma}
\end{figure}

The contour plot in Figure \ref{fig:ContourGamma} show the estimated marginal posterior of $(\gamma_{1}, \gamma_{2})$, corresponding to the \emph{gender} and the \emph{nationality} effect, for the second largest component of the co-authorship data set in Section \ref{Section:DataSet}. These results have been obtained by 3 chains with 30,000 MCMC iterations.

Figure \ref{fig:ContourGamma} reports a positive expected effect of the nationality ($0.45$) and a negligible expected effect of the gender ($0.03$). It can be noted a much larger variability on the gender effect, suggesting a lack of information about the parameter $\gamma_1$.

\begin{figure}[H]
        \centering
        \begin{subfigure}[b]{0.32\textwidth}
                \includegraphics[width=\textwidth]{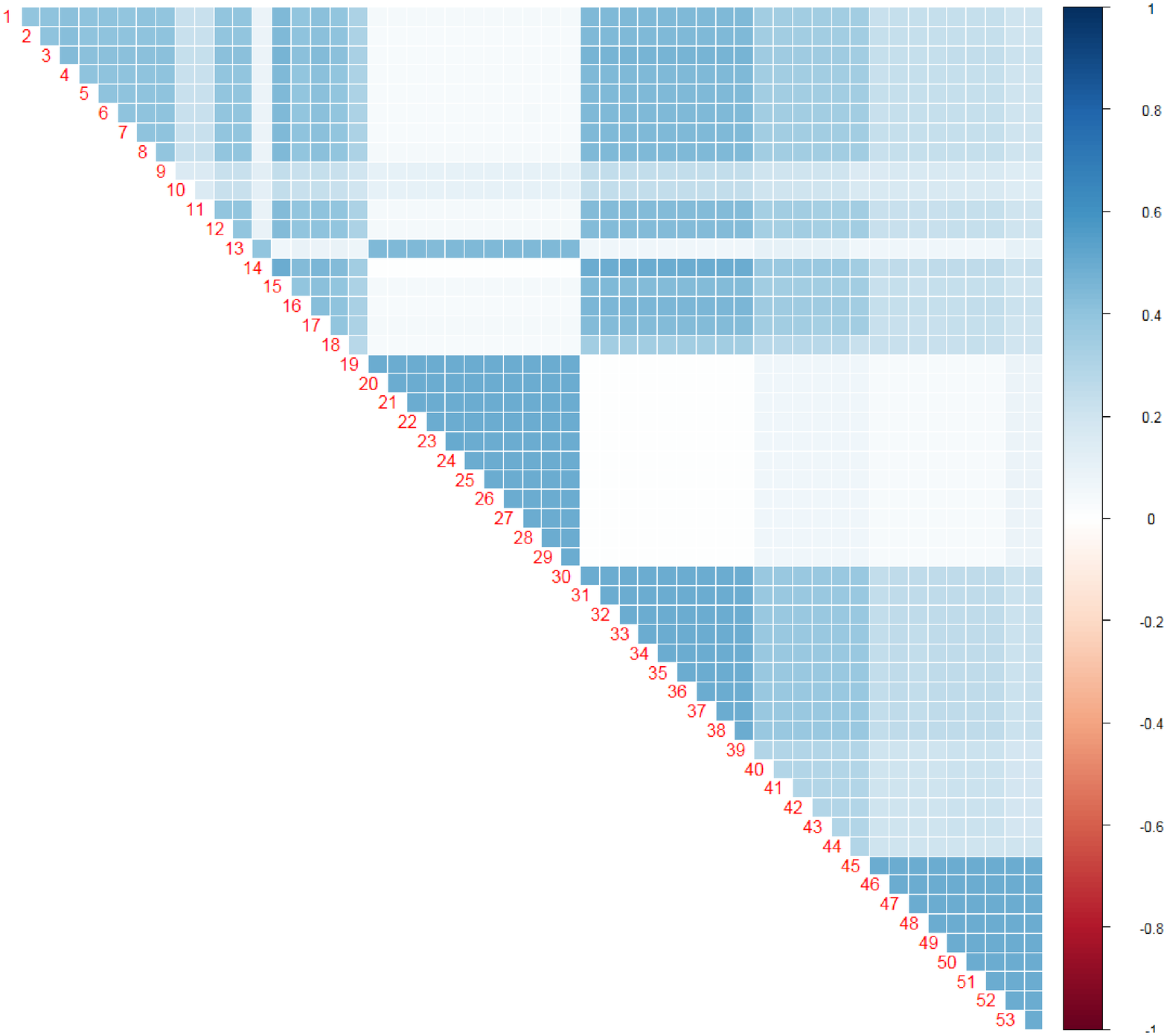}
                \caption{\scriptsize Model (\ref{eq:JointProb_Spec})}
        \end{subfigure}%
        ~
        \begin{subfigure}[b]{0.32\textwidth}
                \includegraphics[width=\textwidth]{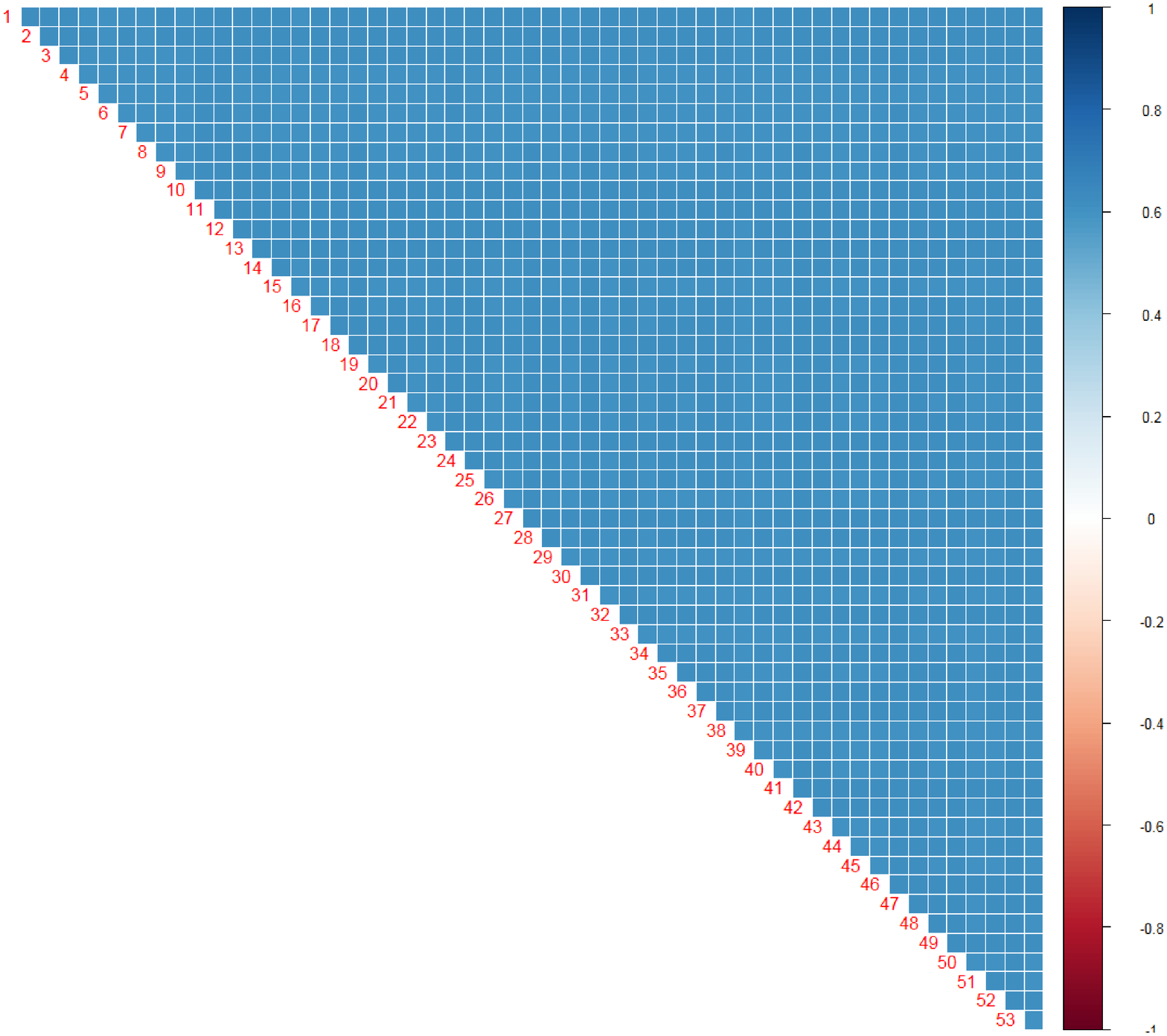}
                \caption{\scriptsize Null model}
        \end{subfigure}
        ~
        \begin{subfigure}[b]{0.32\textwidth}
                \includegraphics[width=\textwidth]{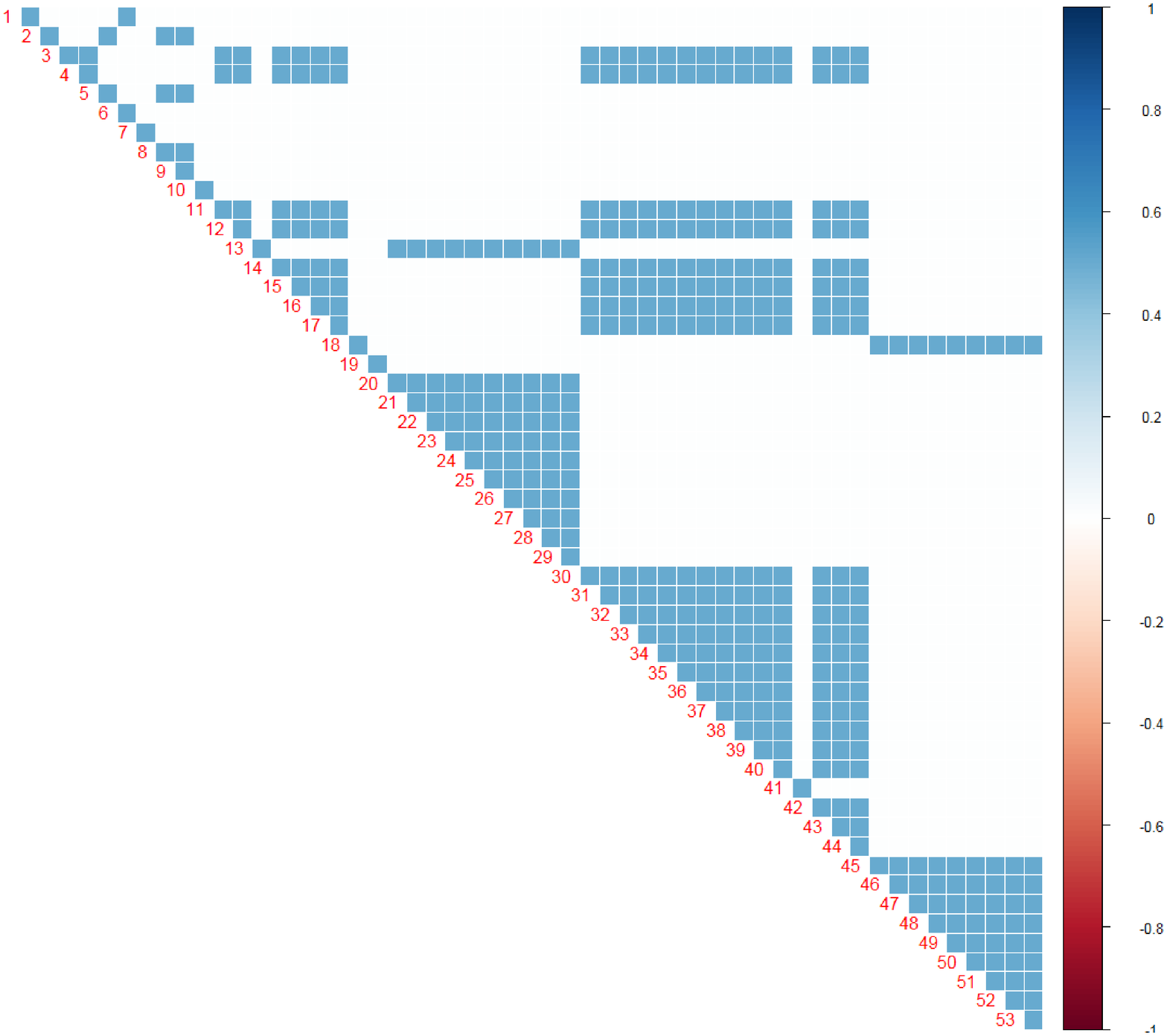}
                \caption{\scriptsize Observed data set}
        \end{subfigure}
\caption{\footnotesize The values of the nationality similarities $\sum_{h_k = 1}^{9} y_{h_k r}y_{h_k s}$ for the $54 \times 53/2$ pairs of nodes .}\label{fig:NodalSimilarity_Nationality}
\end{figure}

After estimating the model parameters $(\boldsymbol{\alpha}, \, \boldsymbol{\gamma})$, a sample of 10,000 elements from $\mathcal{X}$ has been simulated and the nodal similarities have been computed, both for the estimated model and the null model $(\boldsymbol{\alpha}, \, \, \mathbf{0})$. Figure \ref{fig:NodalSimilarity_Nationality} shows the observed and expected nationality similarities for each of the $54 \times 53/2$ pairs of authors. It can be noted that the null model seems to be severely unable to capture the pairwise similarities, whereas the expected values of $\sum_{h_k = 1}^{m_k} y_{h_k r}y_{h_k s}$ under the estimated model in \eqref{eq:JointProb_Spec} substantially resemble the observed similarities.

Tables \ref{tab:GenderProportions} and \ref{tab:NationalityProportions} reports the expected proportions of edges associated to different combinations of genders and nationalities respectively. The estimated model seems to be able to effectively capture the assortative mixing of the data, i.e., the association between nodal properties and connections, along with the total amount of each individual categories and network collaboration density. This was indeed our initial intention, when a joint model for author's characteristics and collaboration pattern was introduced in Section \ref{Section:model}.

\begin{table}[H]
\begin{center}
\scalebox{0.80}{
\begin{tabular}{|r|cc|l|}
\hline
            & male  & female & total  \\
\hline
male         & 0.40 (0.48)  & 0.46 (0.40)  & 0.86 (0.88)  \\
female       & -- --        & 0.14 (0.12)  &   \\
\hline
total &              & 0.60 (0.52)      &\\
            \hline
\end{tabular}}
\caption{\label{tab:GenderProportions} \footnotesize Expected proportions of edged for each combination of genders. The observed proportions are reported within parenthesis. }
\end{center}
\end{table}

\begin{table}[H]
\begin{center}
\scalebox{0.69}{
\begin{tabular}{|r|ccccccccc|}
\hline
             & Italy         & USA           & Spain         & Sweden         & S. Africa     & Japan         & Serbia        & Russia        & UK      \\
\hline
Italy        & 0.004 (0.004) & 0.004 (0.000) & 0.014 (0.017) & 0.001 (0.025) & 0.000 (0.008) & 0.007 (0.000) & 0.000 (0.000) & 0.000 (0.000) & 0.000 (0.000) \\
USA          &               & 0.260 (0.235) & 0.120 (0.031) & 0.012 (0.000) & 0.006 (0.000) & 0.060 (0.004) & 0.005 (0.000) & 0.000 (0.000) & 0.005 (0.023) \\
Spain        &               &               & 0.267 (0.329) & 0.020 (0.024) & 0.010 (0.009) & 0.111 (0.064) & 0.010 (0.201) & 0.000 (0.000) & 0.010 (0.000) \\
Sweden       &               &               &               & 0.001 (0.017) & 0.001 (0.006) & 0.010 (0.000) & 0.000 (0.001) & 0.000 (0.008) & 0.000 (0.000) \\
S.Africa     &               &               &               &               & 0.001 (0.006) & 0.002 (0.000) & 0.000 (0.000) & 0.000 (0.000) & 0.000 (0.000) \\
Japan        &               &               &               &               &               & 0.085 (0.140) & 0.000 (0.000) & 0.000 (0.000) & 0.000 (0.000) \\
Serbia       &               &               &               &               &               &               & 0.000 (0.000) & 0.000 (0.000) & 0.000 (0.000) \\
Russia       &               &               &               &               &               &               &               & 0.000 (0.000) & 0.000 (0.000) \\
UK           &               &               &               &               &               &               &               &               & 0.000 (0.000) \\
\hline
\end{tabular}}
\caption{\label{tab:NationalityProportions} \footnotesize Expected proportions of edged for each combination of nationalities. The observed proportions are reported within parenthesis. }
\end{center}
\end{table}

\begin{figure}[H]
        \centering
        \includegraphics[width=0.60\textwidth]{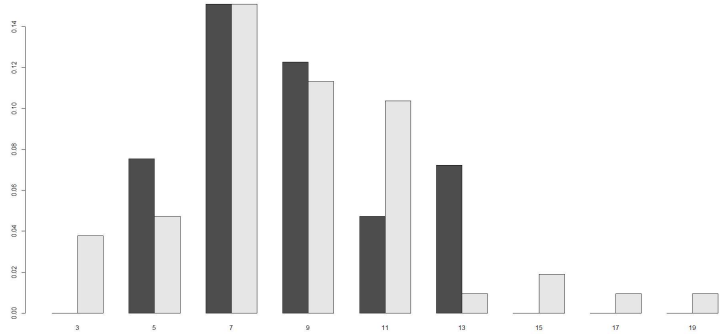}
\caption{\footnotesize In light-grey the observed degree sequence. In dark-grey the expected degree sequence.}\label{fig:DegreeSequence}
\end{figure}

In Figure \ref{fig:DegreeSequence} a graphical comparison between the expected (dark-grey) and the observed (light-grey) degree sequences is provided.

To summarize, for the second largest component of the co-authorship data set, the estimated results confirm the null effect of the gender similarity on the author's connections, along with a positive effect of their nationalities. The estimated model seems to properly fit the empirical observation, as suggested by table \ref{tab:GenderProportions} and \ref{tab:NationalityProportions}. The expected degree sequence also resembles the observed one, suggesting the ability of the model to capture both individual and structural properties of the co-authorship data.

\section{Conclusion}\
\label{Section:Conclusion}
This paper presented an exponential random model for author's characteristics and collaboration pattern in bibliometric networks, which allowed to combine the analysis of multivariate data set with the one of the assortative pattern of nodal similarities in networks. We proposed a Bayesian estimation framework and a specialized MCMC algorithm to simulate from a ``doubly intractable" posterior distribution. We showed a strong capability of the model to account for relevant network features (the degree sequence) based on the observed nodal properties, providing a deep understanding of the linkage between individual and social properties and a substantial insight into the level of \emph{homophily} in co-authorship networks.

Our results suggest several lines of work for future research. We could compare the model fit for different specifications of the sample space $\mathcal{X}$. We could also study the inclusion of further nodal properties, such as age, principal keywords or number of received citations.

\bibliographystyle{wsc}
\bibliography{references}

\end{document}